\documentclass[workingdraft]{usetex-v1}
\usepackage{graphicx}
\usepackage{cite}	
\usepackage{url}

\long\def\com#1{}

\long\def\xxx#1{}

\title{Deterministic Consistency: \\
A Programming Model for Shared Memory Parallelism}
\author{Amittai Aviram and Bryan Ford \\ Yale University}

\begin{document}

\maketitle

\begin{abstract}

The difficulty of developing reliable parallel software
is generating interest in {\em deterministic} environments,
where a given program and input can yield only one possible result.
Languages or type systems can enforce determinism in new code,
and 
runtime systems can impose synthetic schedules on legacy parallel code.
To parallelize existing serial code, however,
we would like a programming model that is naturally deterministic
without language restrictions or artificial scheduling.
We propose {\em deterministic consistency},
a parallel programming model
as easy to understand as
the ``parallel assignment'' construct
in sequential languages such as Perl and JavaScript,
where concurrent threads always read their inputs
before writing shared outputs.
DC supports
common data- and task-parallel synchronization abstractions
such as fork/join and barriers,
as well as non-hierarchical structures
such as producer/consumer pipelines and futures.
A preliminary prototype suggests that software-only implementations of DC
can run applications written for popular parallel environments such as OpenMP
with low ($<10\%$) overhead for some applications.

\end{abstract}

\section{Introduction}
\label{sec-intro}

For decades,
the ``gold standard'' in multiprocessor programming models has been
sequentially consistent shared memory~\cite{lamport79multi}
with mutual exclusion~\cite{hoare72theory}.
Alternative models,
such as explicit message passing~\cite{mpi09}
or weaker consistency~\cite{gharachorloo90memory},
usually represent compromises to improve performance
without giving up ``too much'' of the simplicity and convenience
of sequentially consistent shared memory.
But are sequential consistency and mutual exclusion
really either {\em simple} or {\em convenient}?

In this model,
we find that slight concurrency errors
yield subtle heisenbugs~\cite{lee06problem,lu08learning}
and security vulnerabilities~\cite{watson07exploiting}.
Data race detection~\cite{engler03racerx,musuvathi08heisenbugs}
or transactional memory~\cite{herlihy93transactional,shavit97software}
can help ensure mutual exclusion,
but even ``race-free'' programs may have heisenbugs~\cite{artho03high}.
Heisenbugs result from {\em nondeterminism in general},
a realization that has inspired new languages
that ensure determinism through
communication constraints~\cite{tardieu06scheduling}
or type systems~\cite{bocchino09parallel}.
But to parallelize
the vast body of sequential code
for new multicore systems,
we would like a programming model that is
simple, convenient, deterministic,
and compatible with existing languages.

To this end,
we propose a new memory model
called {\em deterministic consistency} or DC.
In DC, concurrent threads logically share an address space
but never see each others' writes,
except when they synchronize explicitly and deterministically.
To illustrate DC,
consider the ``parallel assignment'' operator
in many sequential languages such as
Python, Perl, Ruby, and JavaScript,
with which one may swap two variables as follows:

\begin{center}
\verb|x,y := y,x|
\end{center}

This construct implies no actual parallel execution:
the statement merely evaluates all right-side expressions
(in some order)
before writing their results to the left-side variables.
Now consider a ``truly parallel'' analog,
using Hoare's notation for fork/join parallelism~\cite{hoare72theory}:

\begin{center}
\verb|{x := y} // {y := x}|
\end{center}

\begin{figure}[t]
\centering
\includegraphics[width=0.45\textwidth]{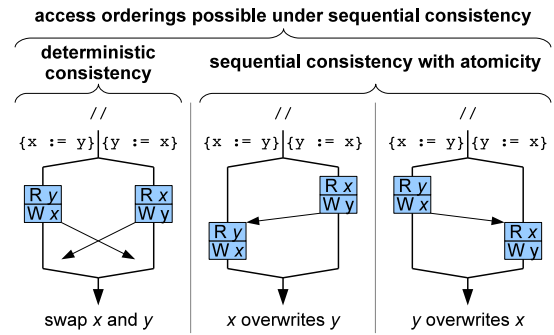}
\caption{Deterministic versus sequential consistency}
\label{fig-swap}
\end{figure}

This statement forks two threads,
each of which reads one variable and then writes the other;
the threads then synchronize and rejoin.
As Figure~\ref{fig-swap} illustrates,
under sequential consistency,
this parallel statement may swap the variables
or overwrite one with the other, depending on timing.
Making each thread's actions atomic,
by enclosing the assignments in critical sections or transactions,
eliminates the swapping case
but leaves a nondeterministic choice
between \verb|x| overwriting \verb|y| and \verb|y| overwriting \verb|x|.
How popular would the former ``parallel assignment'' construct be
if it behaved in this way?
Deterministic consistency, in contrast,
reliably behaves like a parallel assignment:
each thread reads all inputs before writing any shared results.

Like release consistency~\cite{gharachorloo90memory},
DC distinguishes \textit{ordinary} reads and writes from 
\textit{synchronization} operations
and classifies the latter into 
{\em acquires} and {\em releases},
which determine at what point
one thread sees (acquires)
results produced (released) by another thread.
DC ensures determinism by requiring that
(1) program logic uniquely pairs each acquire
	with a matching release,
(2) only an intervening acquire/release pair
	makes one thread's writes visible to another thread, and
(3) acquires handle conflicting writes deterministically.
Unlike most memory models,
reads never conflict with writes in DC:
the swapping example above contains no data race.
A natural way to understand DC---and one way to implement it---%
is as a distributed shared memory~\cite{keleher92lazy,amza96treadmarks}
in which a release explicitly ``transmits''
a message containing memory updates,
and the matching acquire operation ``receives''
and integrates these updates locally.

DC supports not only block-structured synchronization abstractions
such as the fork/join, barrier,
and task constructs of OpenMP~\cite{openmp},
but also non-hierarchical synchronization patterns
such as dynamic producer/consumer graphs and inter-thread queues.
DC can emulate nondeterministic synchronization constructs
in existing parallel code
via techniques such as deterministic scheduling~\cite{
	devietti09dmp,berger09grace,bergan10coredet},
but for new or newly parallelized code, we develop
deterministic alternatives
for common idioms such as pipelines and futures.
A prototype in progress promises to be flexible and efficient enough
for a variety of parallel applications.

\com{
The only common synchronization abstractions we know
that DC does {\em not} directly support
are semantically nondeterministic abstractions
such as mutex locks.

Can we give up
such a common abstraction as mutexes?
For some existing parallel code, probably not.
In this case,
replay~\cite{
	feldman88igor,choi98deterministic,dunlap08execution}
or deterministic scheduling~\cite{
	devietti09dmp,berger09grace,bergan10coredet}
can make any particular execution
of a nondeterministic program exactly reproducible,
by logging or synthesizing an artificial schedule
of the program's nondeterministic events.
But rerunning a program with even slightly different inputs
may change its schedule,
causing hidden schedule-dependent bugs to appear (albeit reproducibly).
Since the vast majority of legacy code is {\em not yet} parallelized,
in confronting the task of parallelizing it,
we would like a programming model
that preserves the {\em natural determinism}
we enjoyed in the uniprocessor world:
a programming model whose abstractions are deterministic
without resort to artificial or accidentally perturbable execution schedules.
We propose DC as one such programming model.
}

Section~\ref{sec-dc} defines DC at a low level, and
Section~\ref{sec-high} explores its use in
high-level environments like OpenMP\@.
Section~\ref{sec-impl} outlines implementation issues,
Section~\ref{sec-related} discusses related work, and
Section~\ref{sec-concl} concludes.

\section{Deterministic Consistency}
\label{sec-dc}

Since others have eloquently
made the case for deterministic parallelism~\cite{
	lee06problem,bocchino09parallel},
we will take its desirability for granted
and focus on deterministic consistency (DC).
This section defines the basic DC model
and its low-level synchronization primitives,
leaving the model's mapping to high-level abstractions
to the next section.

\subsection{Defining Deterministic Consistency}
\label{sec-dc-defn}

As in release consistency (RC)~\cite{
	gharachorloo90memory,keleher92lazy},
DC separates normal data accesses from synchronization operations
and classifies the latter into {\em release},
where a thread makes recent state changes
available for use by other threads,
and {\em acquire},
where a thread obtains state changes
made by other threads.
A thread performs a release when forking a child thread or leaving a barrier,
for example,
and an acquire when joining with a child or entering a barrier.
As in RC,
synchronization operations in DC
are sequentially consistent relative to each other,
and these synchronization operations determine
when a normal write in one thread must become visible to
a normal read in another thread:
namely, when an intervening chain of acquire/release pairs
connects the two accesses in a ``happens-before'' synchronization relation.

\com{
Figure~\ref{fig-ordering}
illustrates four example ordering scenarios involving two threads,
where the horizontal/diagonal arrows indicate 
when a write by one thread becomes visible to the other:
only scenarios (a) and (b) are sequentially consistent,
but all four are permitted under release consistency.
}

While RC relaxes
the constraints of sequential consistency~\cite{lamport79multi},
allowing an even wider range of nondeterministic orderings,
DC in turn tightens RC's constraints
to permit only one unique execution behavior
for a given parallel program.
DC ensures determinism
by adding three new constraints to those of RC:
\begin{enumerate}
\item	Program logic must uniquely pair release and acquire operations,
	so that each release ``transmits'' updates
	to a specific acquire in another thread.
\item	One thread's writes never become visible
	to another thread's reads
	until mandated by synchronization:
	i.e., writes propagate ``as slowly as possible.''
\item	If two threads perform conflicting writes to the same location,
        the implementation handles the conflict deterministically
	at the relevant acquire.
\end{enumerate}

\begin{figure}[t]
\centering
\includegraphics[width=0.40\textwidth]{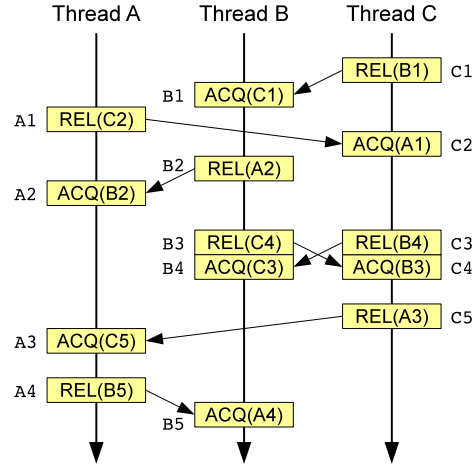}
\caption{Example synchronization trace for three threads
	with labeled and matched release/acquire pairs}
\label{fig-matching}
\end{figure}

Constraint 1 makes synchronization deterministic
by ensuring that a release in one thread
always interacts with the same acquire in some other thread,
at the same point in each thread's execution,
regardless of execution speeds.
A program might in theory satisfy this constraint
by specifying each synchronization operation's ``partner''
explicitly through a labeling scheme.
If each thread has a unique identifier $T$,
and we assign each of $T$'s synchronization actions
a consecutive integer $N$,
then a $(T,N)$ pair uniquely names any synchronization event
in a program's execution.
The program then invokes synchronization primitives
of the form {\tt acquire($T_r,N_r$)} and {\tt release($T_a,N_a$)},
where $(T_r,N_r)$ names the {\tt acquire}'s partner {\tt release}
and vice versa.
Figure~\ref{fig-matching} illustrates a 3-thread execution trace
with matched and labeled acquire/release pairs.
We suggest this scheme only to clarify DC:
explicit labeling would be
an unwelcome practical burden, and
Section~\ref{sec-high} discusses
more convenient high-level abstractions.

\com{
XXX Proposed revision of above:
Abstractly speaking, 
given unique thread identifier $T$ and consecutive integer $N$
to name each synchronization event distinctly, the program
would pair one thread $T_r$'s release $N$ with another
thread $T_a$'s acquire to form quadruple $(T_r,N_r,T_a,N_a)$.
In the next section, we will present ways to map such
release/acquire pairs implicitly to more convenient high-level
abstractions.
}

\com{
This constraint makes acquire/release pairs in DC
quite analogous to the communication channels
modeled in formal process calculi~\cite{XXX},
and some of the reasoning techniques used in these calculi
may apply to DC.

If the relation has cycles, we call the program {\em deadlocked}.
Otherwise, the transitive closure of the relation
forms a partial order on synchronization operations.
}

Constraint 2 makes normal accesses deterministic
by ensuring that writes in a given thread
become visible to reads in another thread at only one possible moment.
Release consistency already requires
a write by thread $T_1$ to become visible to thread $T_2$
{\em no later} than the moment $T_2$ performs an acquire
directly or indirectly following $T_1$'s next release
after the write.
RC permits the write to become visible to $T_2$
{\em before} this point,
but DC requires the write to propagate to $T_2$
at {\em exactly} this point.
By delaying writes ``as long as possible,''
DC ensures that non-conflicting normal accesses behave deterministically
while preserving the key property that makes RC efficient:
it keeps parallel execution as independent as possible
subject to synchronization constraints.

\com{	now mostly (but unfortunately not completely) subsumed
	by the figure in the intro...
\begin{figure*}[t]
\centering
\includegraphics[width=0.90\textwidth]{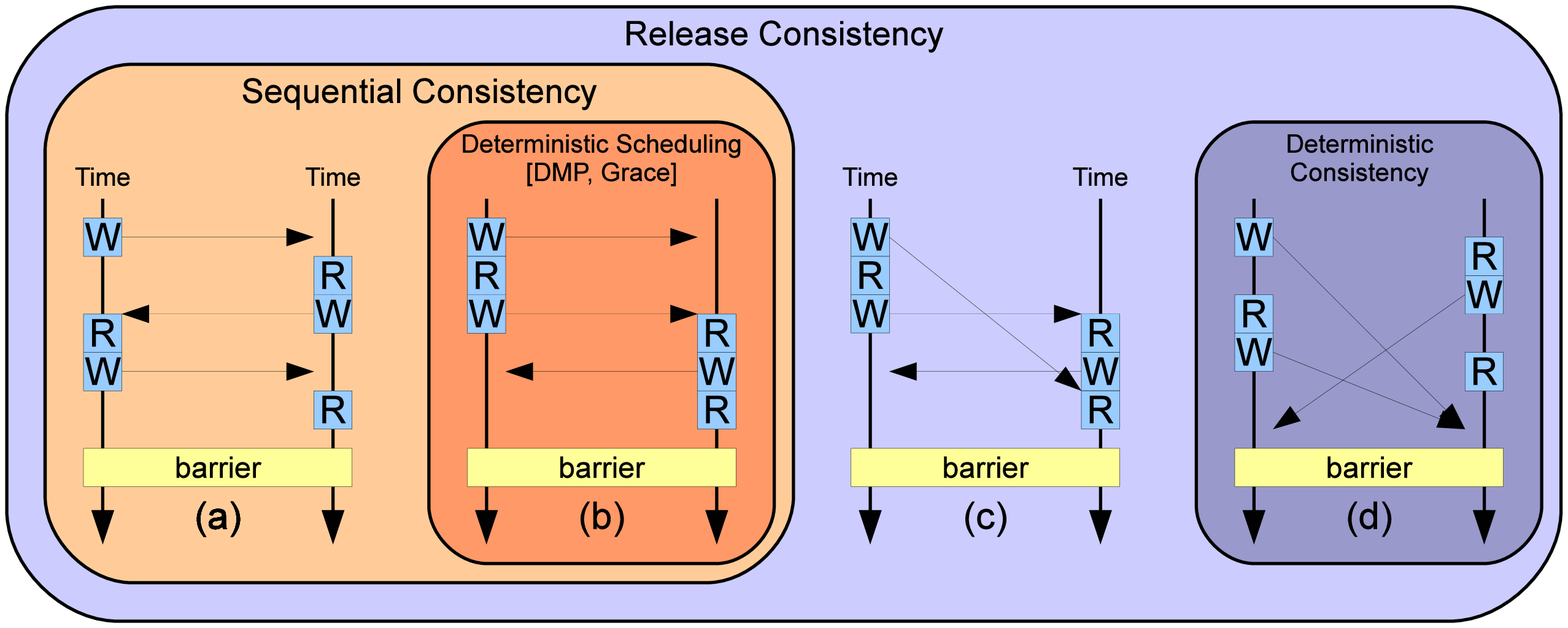}
\caption{Example Event Orderings in Several Consistency Models.}
\label{fig-ordering}
\end{figure*}

Figure~\ref{fig-ordering}
illustrates four possible event orderings
in an example scenerio,
where two threads perform several reads and writes,
then synchronize with a barrier.
Horizontal/diagonal arrows indicate when one thread's writes
become visible to the other thread's reads.
Release consistency permits all four orderings,
while sequential consistency allows only (a) and (b).
Case (b) represents a particular sequential ordering
that might be ``synthesized'' by a deterministic scheduler~\cite{
	devietti09dmp,berger09grace}.
Case (d) represents the unique ordering that DC demands.
}

DC's third constraint affects only programs with data races.
If both threads in Figure~\ref{fig-swap}
wrote to the {\em same} variable before rejoining,
for example,
DC requires the join
to handle this race deterministically.
Since data races usually indicate software bugs,
one response is to throw a runtime exception.
\com{when the memory system detects such a race.}
Other behaviors,
e.g., prioritizing one write over the other,
would not affect correct programs
but may be less helpful with buggy code.

\subsection{Why DC is Deterministic}

To clarify why the above rules adequately
ensure deterministic execution in spite of arbitrary parallelism,
we briefly sketch a proof of DC's determinism.

{\bf Theorem:}
A parallel program whose sequential fragments execute deterministically,
and whose memory access and synchronization behavior
conforms to the rules in Section~\ref{sec-dc-defn},
yields at most one possible result.

{\bf Proof Sketch:}
Assume each synchronization operation explicitly names its ``partner''
as described above.
Suppose we implement DC by accumulating memory ``diffs''
and passing them at synchronization points
atop a message-passing substrate,
as in distributed shared memory~\cite{keleher92lazy,amza96treadmarks}.
Assume the substrate provides
an unlimited number of buffered message channels,
each with a unique name of the form $(T_r,N_r,T_a,N_a)$.
When a thread $T_r$ invokes a {\tt release($T_a,N_a$)} operation
labeled $(T_r,N_r)$,
$T_r$ sends all diffs it has accumulated so far
on channel $(T_r,N_r,T_a,N_a)$.
Similarly, when thread $T_a$ invokes an {\tt acquire($T_r,N_r$)} operation
labeled $(T_a,N_a)$,
it receives a set of diffs on channel $(T_r,N_r,T_a,N_a)$
and applies those it does not already have.
Since each channel $(T_r,N_r,T_a,N_a)$
is used by only one sender $T_r$ and one receiver $T_a$,
the resulting system forms
a Kahn process network~\cite{kahn74semantics},
and DC's determinism follows from that of Kahn networks.

\section{High-level Synchronization}
\label{sec-high}

We are developing {\em DOMP},
a variant of OpenMP~\cite{openmp}
with deterministic consistency.
\com{
OpenMP is a popular high-level parallel environment
standardized for Fortran and C/C++
and also ported to Java~\cite{bull00jomp}.
}
DOMP retains OpenMP's language neutrality and convenience,
supporting most OpenMP constructs
except for fundamentally nondeterministic ones,
and extending OpenMP
to support general reductions
and non-hierarchical dependency structures.

\com{
While most of OpenMP's ``workhorse'' parallel synchronization constructs,
such as parallel {\tt for} loops and barriers,
naturally satisfy DC's constraints as specified above,
we omit a few of OpenMP's semantically nondeterministic constructs
based on mutual exclusion,
and also introduce a few novel, experimental constructs
enabling parallel programs to express
more free-form or dynamic synchronization structures
that OpenMP's lexical block structure does not currently support
without resort to lower-level synchronization primitives.
}

\begin{figure*}[t]
\centering
\includegraphics[width=0.95\textwidth]{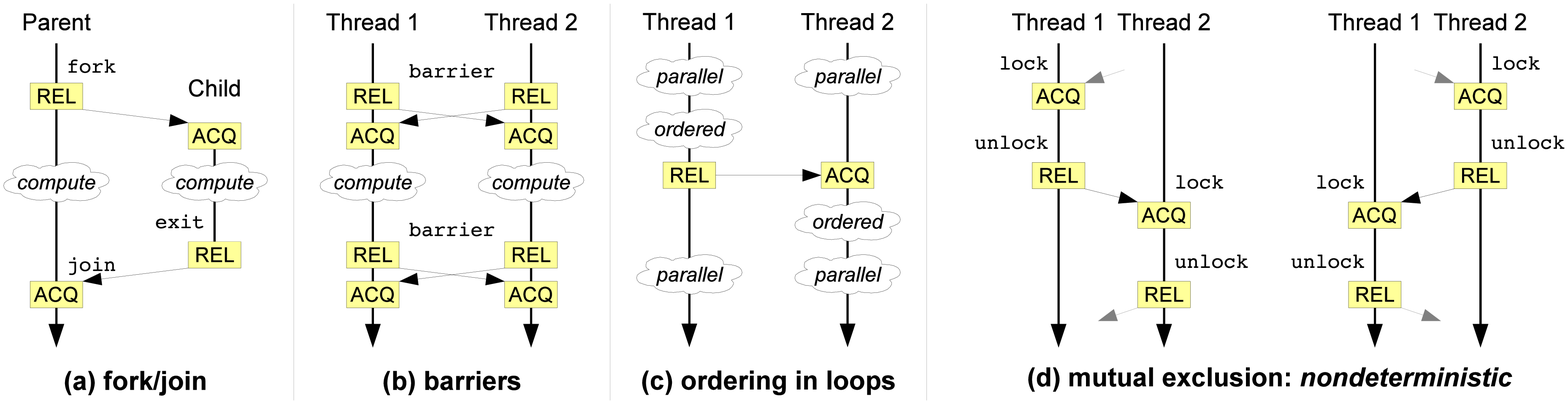}
\caption{Mapping of High-level Synchronization Operations
		to Acquire/Release Pairs}
\label{fig-syncops}
\end{figure*}

\paragraph{Fork/Join:}

OpenMP's foundation is its {\tt parallel} construct,
which forks multiple threads to execute a parallel code block
and then rejoins them.
Fork/join parallelism
maps readily to DC,
as shown in Figure~\ref{fig-syncops}(a):
on fork, the parent releases to an acquire at the birth of each child;
on join, the parent acquires the final results each child releases
at its death.
OpenMP's work-sharing constructs, such as parallel {\tt for} loops,
merely affect each child thread's actions within this fork/join model.

\paragraph{Barrier:}

At a barrier, each thread releases to each other thread,
then acquires from each other thread,
as in Figure~\ref{fig-syncops}(b).
Although we view an $n$-thread barrier
as $n-1$ releases and acquires per thread,
DOMP avoids this $n^2$ cost
using ``broadcast'' release/acquire primitives,
which are consistent with DC
as long as each release matches a well-defined {\em set} of acquires
and vice versa.

\com{	Amittai, I think you misunderstood the semantics of flush -
	e.g., see the flush examples in the OpenMP spec. -baf
Since DC requires threads leaving a barrier to acquire the
new shared state, but ensures their independence between synchronization
events, there is no difference in meaning between
OpenMP's \textit{flush} and \textit{barrier} constructs:  a DC-compliant
implementation of \textit{flush} requires that all threads synchronize
to release and then acquire state.
}

\paragraph{Ordering:}

OpenMP's {\tt ordered} construct
orders a particular code block within a loop by iteration
while permitting parallelism in other parts.
DOMP implements this construct
using a chain of acquire/release pairs among worker threads,
as shown in Figure~\ref{fig-syncops}(c).

\paragraph{Reductions:}

OpenMP's {\tt reduction} attributes and {\tt atomic} constructs
enable programs to accumulate sums, maxima,
or bit masks efficiently across threads.
OpenMP unfortunately supports reductions only on simple scalar types,
leading programmers to serialize complex reductions unnecessarily
via {\tt ordered} or {\tt critical} sections or locks.
All uses of these serialization constructs
in the NAS Parallel Benchmarks~\cite{jin99openmp}
implement reductions, for example.
DOMP therefore provides a generalized {\tt reduction} construct,
by which a program can specify a custom reduction
on pairs of variables of any matching types, as in this example:

\begin{small}
\begin{verbatim}
#pragma omp reduction(a:a1,b:b1,c:c1)
  { a += a1; b = max(b,b1);
    if (c1.score > c.score) c = c1; }
\end{verbatim}
\end{small}

DOMP accumulates each thread's partial results in thread-private variables
and reduces them at the next join or barriar via combining trees,
improving both convenience and scalability over serialized reduction.

\paragraph{Tasks:}

OpenMP 3.0's {\tt task} constructs express a form of fork/join parallelism
suited to dynamic work structures.
Since DC rules prevent a task from seeing any writes of other tasks
until it completes and synchronizes at a barrier or {\tt taskwait},
DOMP eliminates OpenMP's risk of subtle bugs
if one task uses shared inputs
that are freed or go out of scope in a concurrent task.

\com{	this is just specification nondeterminism, which is nonessential. -baf
OpenMP 3.0 includes a new \textit{task} construct, which its specification
leaves open to nondeterminism, since any encountering thread may execute
the designated task.  A DC-compliant implement must impose a scheduling order
that ensures that the same pair of threads will always meet at the release-acquire
that implicitly comes right before the task.  In practice, this scheduling constraint
should not impose any special burden on OpenMP programs, since any implementation
already assigns tasks to threads according to some algorithm.  We would only guarantee
that this algorithm is itself deterministic.
}

DOMP extends OpenMP with explicit {\em task objects},
with which a {\tt taskwait} construct can name and synchronize
with a particular {\tt task} instance independently of other tasks,
in order to express
futures~\cite{halstead85multilisp}
or non-hierarchical dependency graphs~\cite{edwards08programming}
deterministically:

\begin{small}
\begin{verbatim}
omp_task mytask;
#pragma omp task(mytask)
  { ...task code... }
...other tasks...
#pragma omp taskwait(mytask)
\end{verbatim}
\end{small}

\com{
\paragraph{Task Group}

The new \textit{task} construct offers us an opportunity to extend the OpenMP API
in order to accommodate a more free-form, nonhierarchical style of parallelism that
still ensures deterministic execution, as we see, for instance, in the SHIM
implementation of a jpeg decoder \cite{edwards08programming}.
To that end, we define a new \textit{taskgroup}
construct, which encloses any number of \textit{task} constructs.  At the close of
the \textit{taskgroup} construct, the state changes which the threads executing its
tasks have wrought remains available for acquisition at any time dictated by program logic,
and expressed explicitly with a new \textit{getresults} construct.
\begin{verbatim}
#pragma omp taskgroup foo
{
  #pragma omp task 
  {
    a = f(x);
  }
  #pragma omp task
  {
    b = g(y);
  }
}

c = h(w);

#pragma omp getresults foo

d = k(a, b);
\end{verbatim}

Before \textit{getresults},
no threads see any of these state changes, except each thread performing its respective
task, which only sees its own changes.  In this way, the pairing of the
collection of threads in a \textit{taskgroup}
with the parent thread at \textit{getresults} works 
like two nodes at either end of a communication channel in
a Kahn process network.  

}

\paragraph{Mutual exclusion:}

Unlike {\tt ordered},
which specifies a {\em particular} sequential ordering,
mutual exclusion facilities
such as {\tt critical} sections and locks
imply an arbitrary, {\em nondeterministic} ordering.
Mutual exclusion violates Constraint 1 in Section~\ref{sec-dc-defn}
because it permits multiple acquire/release pairings,
as illustrated in Figure~\ref{fig-syncops}(d).
While DOMP could emulate mutual exclusion via deterministic scheduling,
we prefer to focus on developing deterministic abstractions
to replace common uses of mutual exclusion,
such as general reductions.

\paragraph{Flush:}

Some OpenMP programs implement custom synchronization structures
such as pipelines
using the {\tt flush} (memory barrier) construct in spin loops.
Like mutual exclusion, DOMP omits support for such constructions,
in favor of expressing dependency graphs such as pipelines
deterministically using task objects.

\section{Implementing DC}
\label{sec-impl}

We have built an early user space prototype
implementing DC with a pthreads-like fork/join API.
The prototype encouragingly shows less than $10\%$ overhead
on the coarse-grained PARSEC benchmarks~\cite{bienia08characterization}
Blackscholes and Swaptions.
Finer-grained benchmarks such as Streamcluster
currently show high overheads,
but many optimization opportunities remain.
The rest of this section outlines key challenges and opportunities
in implementing deterministic consistency,
for both shared memory multithreaded programs and multiprocess systems.

\subsection{Shared Memory Challenges}

\paragraph{Memory Access Isolation:}
Since DC requires one thread's writes
to remain invisible to a second thread
until the two threads synchronize,
the threads must effectively execute
in separate ``workspaces'' between synchronization events.
Virtual memory and write-sharing techniques like those used
to implement lazy release consistent
distributed shared memory~\cite{amza96treadmarks}
should apply to DC.
Memory accesses may also be isolated
via instruction-level rewriting~\cite{bergan10coredet},
possibly reducing the cost of synchronization operations
at the expense of adding overhead to all ordinary memory accesses.
Hardware support~\cite{gharachorloo90memory,devietti09dmp}
could mitigate the performance cost of isolation,
but is unlikely to appear in commodity hardware
unless software-based approaches first demonstrate
deterministic parallelism to be viable and compelling.

\com{	moved from sec 3...
The shared variables to which concurrent threads write new values are 
``ordinary'' rather than ``synchronization'' variables.  Our implementation
provides each newly-spawned thread with its own private logical copy
of all shared variables, to which the threads may then write new values,
and, at the close of the block, the implementation reconciles all changes
into the resulting new state, issuing a data race exception if any two threads 
have changed the value of the same variable.  
}

\paragraph{Shared Resources:}
Shared resources in current environments
implicitly introduce nondeterminism through mutual exclusion:
calling {\tt malloc()} concurrently in multiple threads
may yield different pointers depending on execution timing,
for example,
and the file descriptor number returned
by a call to Unix's {\tt open()}
may have similar timing dependencies
on other threads' file descriptor operations.
The {\tt malloc()} problem may be addressed
by assigning each thread a separate virtual memory address range
and allocation pool from which to satisfy {\tt malloc()} requests;
such an allocator may also benefit scalability.
The file descriptor table problem might be addressed
by using higher-level equivalents such as {\tt fopen()} that
do not imply mutual exclusion.
These approaches do not address shared resources
outside the application process, however,
such as reads and writes to shared files in an external file system.

\subsection{Beyond Shared Memory}

While we have focused on the intra-process shared memory abstraction,
DC may also be applicable at the system level
for state shared among processes.
Standard operating systems,
for example,
commonly give all processes sequentially consistent access
to a globally shared file system
(though network file systems often relax consistency somewhat).
This design yields the same problems of nondeterminism and heisenbugs
at inter-process level that we see within multithreaded programs:
we find often that a large software source tree
builds reliably under a sequential `{\tt make}'
but fails nondeterministically under a parallel `{\tt make -j}' command,
for example.

In place of sequential consistency,
an OS might provide
a deterministically consistent file system to processes,
enabling a multi-process computation to run deterministically
even as processes share state by reading and writing files.
If a parallel {\tt make} forks off two compiler instances
running in parallel, for example,
each compiler would execute in its own private virtual copy of the file system
until completion;
the system would then reconcile the {\tt .o} files produced by each compiler
into a single directory once both compilers complete.

There will always be shared resources
``outside the reach'' of any deterministic environment,
whose use will introduce nondeterminism into the program:
for example, I/O requests arriving at a network server from its clients.
In such cases
the only solution may be to accept some nondeterminism,
log nondeterministic inputs to enable later replay,
or avoid their use entirely.

\section{Related Work}
\label{sec-related}

DC conceptually builds on release consistency~\cite{gharachorloo90memory}
and lazy release consistency~\cite{keleher92lazy},
which relax sequential consistency's ordering constraints
to increase the independence of parallel activities.
DC retains these independence benefits,
additionally providing determinism
by delaying the propagation of any thread's writes to other threads
until {\em required} by explicit synchronization.

Race detectors~\cite{engler03racerx,musuvathi08heisenbugs}
can detect certain heisenbugs,
but only determinism eliminates their possibility.
Language extensions
can dynamically check determinism assertions in parallel code~\cite{
	sadowski09singletrack,burnim09asserting},
but heisenbugs may persist if the programmer omits an assertion.
SHIM~\cite{
	edwards06shim,tardieu06scheduling,edwards08programming}
provides a deterministic message-passing programming model,
and DPJ~\cite{bocchino09dpj,bocchino09parallel}
enforces determinism in a parallel shared memory environment
via type system constraints.
While we find language-based solutions promising, 
parallelizing the huge body of existing sequential code
will require parallel programming models compatible with existing languages.

DMP~\cite{devietti09dmp,bergan10coredet}
uses binary rewriting
to execute existing parallel code deterministically,
dividing threads' execution into fixed ``quanta''
and synthesizing an artificial round-robin execution schedule.
Since DMP is effectively
a deterministic {\em implementation}
of a nondeterministic programming model,
slight input changes may still
reveal schedule-dependent bugs.
Grace~\cite{berger09grace}
runs fork/join-style programs deterministically
using virtual memory techniques.
These systems still pursue sequential consistency as an ``ideal''
and rely on speculation for parallelism:
if a thread reads a variable concurrently written by another,
as in the ``swap'' example in Section~\ref{sec-intro},
one thread aborts and re-executes sequentially.
A partial exception is DMP-B~\cite{bergan10coredet},
which weakens consistency within a parallel execution quantum.
DC, in contrast, keeps threads fully independent
between program-defined synchronization points,
never requires speculation or rollback,
and imposes no artificial execution schedules
prone to accidental perturbation.


Replay systems
can log and reproduce particular executions
of conventional nondeterministic programs,
for debugging~\cite
	{curtis82bugnet,leblanc87debugging}
or intrusion analysis~\cite{dunlap02revirt,joshi05detecting}.
The performance and space costs of logging nondeterministic events
usually make replay usable only ``in the lab,'' however:
if a bug or intrusion manifests under deployment with logging disabled,
the event may not be subsequently reproducible.
In a deterministic environment, any event is reproducible
provided only that the original external inputs to the computation are logged.
\com{
Time-travel (replay/reverse-execution) process debugging:
Curtis, "BugNet: a debugging system for parallel programming environments", 1982
Leblanc, "Debugging parallel programs with instant replay", 1987
Pan, "Supporting reverse execution for parallel programs", 1988
Feldman, "Igor: A System for Program Debugging via Reversible Execution", 1988
Narayanasamy, "BugNet", ISCA 2005
Geels, "Replay Debugging for Distributed Applications", USENIX 2006

Time-travel OS debugging:
King, "Debugging operating systems with time-traveling virtual machines", 2005
}


As with deterministic release consistency,
transactional memory (TM) systems~\cite
	{herlihy93transactional,shavit97software}
isolate a thread's memory accesses from visibility to other threads
except at well-defined synchronization points,
namely between transaction start and commit/abort events.
TM offers no deterministic ordering between transactions, however:
like mutex-based synchronization,
transactions guarantee only atomicity, not determinism.
\com{
	Other TM literature:
	McRT-STM: a high performance software transactional memory system
	Dynamic Performance Tuning of Word-Based Software Transactional Memory
	Transactional Memory with Strong Atomicity, PPoPP 2009
}


\com{
Guava - Java dialect:
D. F. Bacon et al. Guava: A dialect of Java without data races. In
Object-Oriented Programming, Systems, Languages, and Applica-
tions (OOPSLA), pp. 382–400, Minneapolis, Minnesota, Oct. 2000.
- not deterministic, actually, only atomic!

Deterministic message passing languages: StreamIt~\cite{XXX}, SHIM~\cite{XXX}

[16] W. Thies, M. Karczmarek, and S. Amarasinghe. StreamIt: A lan-
guage for streaming applications. In Compiler Construction (CC),
volume 2304 of LNCS, pp. 179–196, Grenoble, France, Apr. 2002.

O. Tardieu and S. A. Edwards. Scheduling-independent threads and exceptions in SHIM. In Embedded Software (Emsoft), pp. 142–151,
Seoul, Korea, Oct. 2006.


Speculation: Cilk
R.D.Blumofeetal.Cilk:Anefficientmultithreadedruntimesystem. In Principles and Practice of Parallel Programming (PPoPP), pp.
207–216, Santa Barbara, CA, July 1995.

XXX more from Martin Vechev:
CoreDet
- LLVM-based, still bad base efficiency, but has the basics of DRC

Berger, Grace: Safe Multithreaded Programming for C/C++
- VM-based, but still speculative, pursuing sequential consistency

SingleTrack: A Dynamic Determinism Checker for Multithreaded Programs

Burnim, Asserting and Checking Determinism for Multithreaded Programs
- runtime determinism checker; can deal with FP arithmetic reordering.
}

\section{Conclusion}
\label{sec-concl}

Building reliable software on massively multicore processors
demands a predictable, understandable programming model,
a goal that may require giving up
sequential consistency and mutual exclusion.
Deterministic consistency provides an alternative parallel programming model
as simple as ``parallel assignment,''
and supports existing languages
and synchronization abstractions.

\bibliography{os}
\bibliographystyle{plain}

\end{document}